\documentclass[aps,pre,groupaddress,showpacs,twocolumn]{revtex4}
\usepackage{epsfig}
\usepackage{amsmath}

\begin{document}
\title{Localization transition on complex networks via spectral statistics}
\author{M. Sade, T. Kalisky, S. Havlin and R. Berkovits}
\affiliation{The Minerva Center, Department of Physics,
    Bar-Ilan University, Ramat-Gan 52900, Israel}
\date{Oct. 21, 2005, version 3.1}
\begin{abstract}
  
  The spectral statistics of complex networks are numerically studied.
  The features of the Anderson metal-insulator transition are found to
  be similar for a wide range of different networks.  A
  metal-insulator transition as a function of the disorder can be
  observed for different classes of complex networks for which the
  average connectivity is small. The critical index of the transition
  corresponds to the mean field expectation.  When the connectivity is
  higher, the amount of disorder needed to reach a certain degree of
  localization is proportional to the average connectivity, though a
  precise transition cannot be identified.  The absence of a clear
  transition at high connectivity is probably due to the very compact
  structure of the highly connected networks, resulting in a small
  diameter even for a large number of sites.
\end{abstract}

\maketitle

The Anderson transition predicts a transition from extended
(metallic) to localized (insulating) eigenstates as a function of
the disorder or energy of a quantum system \cite{anderson58}. 
This second-order phase transition turns out to be 
a very general property related to the
transport of quantum and classical
waves in disordered systems and signatures of it have been observed
for electrons in metals, microwaves in waveguides, light in liquids and gels 
and acoustical waves in the earth crust\cite{bergmann84}. 
For all these systems the clearest signature
of the transition is the different spread of a signal injected into the
system. While in a metallic phase the injected wave will spread all over
the system, in the insulating phase it will be localized in the vicinity
of the injection point. This effect is a result of the constructive
interference between time-reversed path throughout the system. Thus,
the details of the transition are strongly influenced by the dimensionality
and topology of the system.

The lower critical
dimension, below which the system is localized for all values of
disorder, is believed to be two \cite{abrahams79}, since the probability of
returning to the origin (i.e., constructive interference due to
time reversal symmetry) is finite below $d=2$. The upper critical
dimension \cite{castellani86} (beyond which the critical exponents reach 
their mean-field values) remains uncertain although it is argued to
be infinity \cite{mirlin94}. The parameters defining the transition are
traditionally given as
the critical disorder $W_c$ expressed in terms of the width of the
distribution from which the on-sites energies in the Anderson model
are drawn, and the
critical exponent $\nu$ (for definitions see Sec. \ref{s2}). 
For square lattices with dimensionality $d=3,4$ the values of $W_c$ and $\nu$
are well established : for $d=3$, $W_c \sim 16.5$ and $\nu \sim
1.5$ \cite{kramer93,slevin00}, while for 
$d=4$, $W_c \sim 35$ and $\nu \sim 1$ \cite{schreiber96,zharekeshev98}.
For higher dimensions the following extrapolation was offered
\cite{schreiber96}: $\nu \sim 0.8/(d-2) + 0.5$  and $W_c \sim
16.5(d-2)$, which was obtained by studying the transition on
bifractal topologies. Thus, the mean field critical index value of
$1/2$ is obtained in the the upper critical dimension $d=\infty$.

There has been recently much interest in the properties of random
scale-free networks \cite{albert00,barabasi,dorogovtsev03,vespignani04,cohen00,cohen03,bollobas02,molloy98,callaway00,newman01}. 
These networks are characterized by the
fact that each node is connected with some finite probability to
any other node in the graph, which is very different from the
usual topology of a real space lattice in which nodes are
connected only to their neighbors. 
This leads to a very interesting
behavior of the graphs when properties such as percolation, 
cluster structures, paths length etc. are considered \cite{barabasi}. 
Interest in the influence this unusual topology has
on the properties of wave interference (i.e., the Anderson
transition of these graphs) is rising. Indeed, recently 
the Anderson transition in particular networks, namely the
small-world networks \cite{zhu00,giraud05} and the Cayley tree 
\cite{sade03} were studied. Older work on the localization
properties of sparse random matrices \cite{fyodorov91} is directly
relevant to Erd\"os-R\'enyi graphs.
Since scale free networks have an unusual
topology for which anomalous classical properties 
have been found \cite{braunstein03,lopez05} it is
of particular interest to study the Anderson localization
in these networks.

Essentially, the probability to return to the origin defines
the dimensionality of a system for the Anderson transition.
Therefore, one may speculate that random graphs, 
which have only very long closed trajectories,
correspond to systems with an infinite
dimension \cite{bollt04}. 
On the other hand, the critical disorder which depends
roughly on the number of nearest neighbors $Z$ is expected to
follow \cite{anderson58} $W_c \sim Z$ , which for a random graph with an
average degree $\langle k \rangle$ 
(i.e., the average number of connections per node)
corresponds to $W_c \sim
\langle k \rangle$. Thus, one may expect here an interesting
situation in which the critical index $\nu$, which is determined
by the dimensionality is close to a half, while the critical
disorder is determined by $\langle k \rangle$. This is very different
than the situation described by the extrapolation given above, where
the critical disorder for an infinite dimension should also
be infinite.

Beyond the general interest in investigating the Anderson transition
on scale free networks, and the fresh outlook it might 
provide on the localization
phenomenon, the metal-insulator transition can provide 
insights into the functionality
of complex networks. Consider for example an optical communication 
network.  In such a network the edges of the graph represent optical fibers 
in which light propagates and the nodes represents a beam
splitter which redistributes the incoming wave into the outgoing
bonds connected to the node. 
Since for high quality optical fiber there are essentially no losses 
or decoherence on the bonds, the amplitude of the transmitted wave
will not depend on the bond length (on the other hand phase 
will depend on the length). This network may be mapped on a
tight binding Hamiltonian of the type described 
in Sec. \ref{s2} \cite{shapiro82}.
An interesting question for such a scale
free network is whether a wave injected into one of the nodes will
produce a signal at all other node (which in the language of the Anderson
localization is equivalent to the question is the system metallic) or
only at a finite set of other nodes (i.e., the system is insulating). 
A metal-insulator transition in this network corresponds to a phase transition 
between those two phases as function of the properties of the nodes.
Generally, for any complex network in which information propagates 
in a wave-like
fashion and interference is possible, the Anderson transition will limit the
spread of the information throughout the network.

The paper is organized in the following way:
In the next section (Sec. \ref{s1}) 
we describe the different networks which were
considered, while in Sec \ref{s2} the spectral statistics method
which was applied in order to identify the metal-insulator transition
is outlined. In Sec. \ref{s3} the results are depicted and some
general characteristics of the localization on complex networks
are discussed.

\section{Characteristics of the different networks}
\label{s1}

Our main goal is to study the Anderson transition for different complex
networks. In this section we shall define the characteristics of the
networks which will be considered.

\subsection{Random Graph}

A random graph (or- random regular graph) is a graph with 
$N$ nodes, each is connected to 
exactly $k$ random neighbors \cite{barabasi}.
The diameter of a graph is the maximal distance between any pair of its nodes.
In a random graph the diameter $d$ is proportional to $\ln N$.
In Sec. \ref{s3} we shall present results of the level spacing distribution
for random-regular graphs with 
$k=3$ .

\subsection{Erd\"os-R\'enyi Graphs}

In their classical model from 1959 Erd\"os and R\'enyi (ER) \cite{erdos59}
describe a graph with $N$ nodes
where every pair of nodes is connected with probability $p$
resulting in $\langle k \rangle=Np$.
For a large random graph the degree distribution follows the 
Poisson distribution:
\begin{equation}
P(k)=e^{-\langle k \rangle}\frac{\langle k \rangle^k}{k!}.
\label{pois}
\end{equation}
The diameter of such a graph follows:
$d \sim \ln {N}$, similar to a random graph.
In Sec. \ref{s3} we have specifically calculated the level distribution
for $\langle k \rangle = 3, 3.1, 3.2, 3.5, 4, 5, 7.5$ and $10$.

\subsection{Scale-Free Networks}

Scale free (SF) networks \cite{albert00}
are networks where the degree distribution 
(i.e., fraction of sites with k connections) decays as a power-law.
The degree distribution is given by \cite{review6} :

\begin{eqnarray}\nonumber
 P(k)=ck^{-\lambda}, m < k < K,
\label{psf}
\end{eqnarray}
where $c=(\lambda-1)m^{\lambda-1}$ and $K=mN^{1/(\lambda-1)}$ \ 
\cite{cohen00}, 
$\lambda$ is the power-law exponent, $m$ is a lower cutoff, 
and $K$ is an upper cutoff . Thus, there are no sites
with degree below $m$ or above $K$.
The diameter of the SF network can be regarded as the mean distance 
of the sites from the site with the
highest degree. For graphs with $2<\lambda<3$ the distance behaves 
as $d \sim \ln(\ln(N))$\ \cite{cohen03}, and for $\lambda=3$ as
$d \sim \ln(N)/\ln(\ln(N))$ \cite{bollobas02}. This anomalous behavior stems from the
structure of the network where a small core 
containing most of the high degree sites has a very
small diameter. For higher values of $\lambda$ the distance behaves 
as in ER, i.e., $d \sim \ln(N)$.
The $\langle k \rangle$ of a SF graph is obtained by the 
following expression:
\begin{equation}
\langle k \rangle=\frac{1-\lambda}{2-\lambda} \times 
\frac{K^ {2-\lambda} -m^{2- \lambda} }{K^ {1-\lambda} -m^{1- \lambda}} 
\label{kavesf}
\end{equation}
For $\lambda>2$ and large enough $N$ the average degree ,$\langle k \rangle$, is a constant.
The SF networks analyzed in this paper correspond to 
$\lambda = 3.5 ,4 ,5$ with $m=2$ (lower cutoff), and
$\lambda = 4 ,5$ with $m=3$. Due to their small diameter, SF
networks with $\lambda<3$  were not considered.

\subsection{Double peaked distributions}

In order to find hierarchical relation between the different graphs, we
studied also some variations on these graphs. For a random
graph we changed the degree of a small percent of the
nodes, so we have a graph with double peaked distribution. Thus,
the average connectivity, $\langle k \rangle$, is the average
degree of the nodes. Several examples were taken: changing 5\% of
the nodes to $k=5$ (instead of $k=3$) resulting in $\langle k
\rangle = 3.1$, or changing 5\% of
the nodes to $k=10$ ($\langle k \rangle = 3.35$).
Replacing 20\% of the nodes connectivity for the previous cases will
result in $\langle k \rangle = 3.4$ (for $k=5$
nodes) and $\langle k \rangle = 4.4$ (for $k=10$ nodes).
Additionally, in order to relate with
previous results of the metal insulator transition on a Cayley-tree \cite{sade03} we
checked a tree in which 5\% of its nodes have higher degree ($k=4$) resulting in
an average connectivity $3.05$ and creating few closed trajectories - loops.

\section{Method}
\label{s2}

Now we turn to the calculation of the spectral statistics of
these networks.
First, one must construct the appropriate network structure, i.e.,
to determine which node is connected to which. This is achieved
using the following algorithm \cite{cohen00,molloy98}:
\\
1. For each site choose a degree from the required distribution.\\
2. Create a list in which each site is repeated as many times as 
its degree.\\
3. Choose randomly two sites from the list and connect this pair of site
as long as they are different sites.\\
4. Remove the pair from the list. Return to 3.\\

The diameter of a graph is calculated by building shells of sites \cite{review6}.
The inner shell contains  the node
with the highest degree, the next contains all of its neighbors, and so on.
Of course, each node is counted only 
once. The diameter of the system is then determined by the
number of shells.
Two more options which were considered are defining the
diameter by the most highly populated shell, or by
averaging over the shells. The diameter obtained by the various
methods are quite similar.

The energy spectrum is calculated using the usual tight-binding Hamiltonian,
\begin{equation}
   H=\sum_{i}\varepsilon_i a_i^{\dag} a_i -
   \sum_{\langle i,j \rangle} a_j^{\dag} a_i , \label{hamiltonian}
\end{equation}
where first term of the Hamiltonian 
stands for the disordered on-site potential on each node $i$ of the
network. 
The on-site energies, $\varepsilon_i$ are uniformly
distributed over the range $-W/2\leq\varepsilon_i\leq W/2 $. 
The second term corresponds to the hopping matrix element which is set to $1$,
and $\langle i,j \rangle$ denotes nearest neighbor nodes which are
determined according to the network structure.
We diagonalize the Hamiltonian exactly, 
and obtain $N$ eigenvalues $E_i$ (where $N$
is the number of nodes in the graph) and eigenvectors $\psi_i$. 
Then we calculate the distribution $P(s)$ of adjacent
level spacings $s$, where 
$s= {(E_{i+1}-E_{i}) } / {\langle E_{i+1}-E_{i} \rangle}$,
and $\langle \ldots \rangle$ denotes averaging over different realizations
of disorder and when relevant also over different realizations of node
connectivities.

\begin{figure}\centering
\epsfxsize6cm\epsfbox{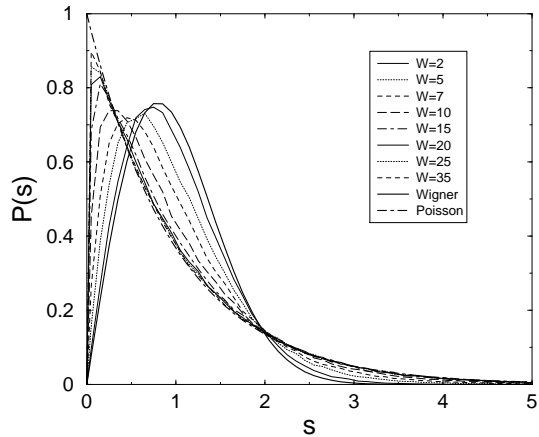} \caption{The distribution
$P(s)$ for a $500$ sites scale-free graph with $\lambda=4$ ($m=2$). A clear transition from Wigner to Poisson is observed
as a function of disorder. } \label{fig1}
\end{figure}

One expects the distribution to shift as function of the on-site disorder
from the Wigner surmise distribution (characteristic of
extended states):
\begin{eqnarray}
   P_W(s)=\frac{\pi s}{2}\exp \left[-\frac{\pi s^2}{4} \right],
\label{wigner}
\end{eqnarray}
at weak disorder to a Poisson distribution (characteristic of localized states)
at strong disorder:
\begin{eqnarray}
   P_P(s)=e^{-s}.
\label{poisson}
\end{eqnarray}
An example for such a transition is presented in Fig. \ref{fig1} where
a scale-free graph with $\lambda=4$ and $m=2$ was considered. 
As $W$ increases $P(s)$
shifts toward the Poisson distribution. Additional hallmark features
of the Anderson transition, such as
the fact that all curves intersects at $s=2$ and the peak of the
distribution "climbs" along the Poisson curve for larger
values of $W$ are also apparent. Similar transition from Wigner
to Poisson statistics is seen also for the other networks considered
in this study.

\begin{figure}\centering
\epsfxsize6cm\epsfbox{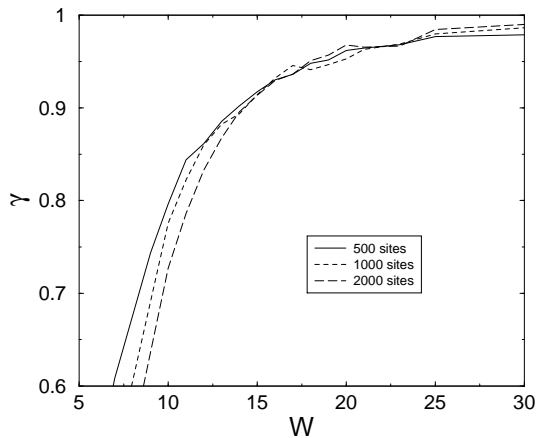} \caption{ $\gamma$ as function
of $W$ for different SF graphs sizes ($\lambda=4$ and $m=2$). The typical behavior for
finite size transition is seen, where a crossing in the size
dependence of $\gamma$ between the metallic (small values of $W$)
and localize (large value of $W$) regime is seen.} \label{fig2}
\end{figure}

The transition point can be determined more accurately 
from calculating \cite{shklov92}
\begin{equation}\label{gamma}
    \gamma = \frac{{\int_{2}^{\infty}P(s) ds}-{\int_{2}^{\infty}P_{w}(s) ds}}
    {{\int_{2}^{\infty}P_{p}(s) ds}-{\int_{2}^{\infty}P_{w}(s) ds}},
\end{equation}
where $\gamma\rightarrow 0$ as the distribution tends toward 
the Wigner distribution, and $\gamma\rightarrow 1$ if 
the distribution approaches the Poisson distribution. 
One expects that as the system size increases, the finite size 
corrections will become smaller resulting in a distribution closer 
to a Wigner distribution in the metallic regime and to 
Poisson in the localized one.
At the transition point the distribution should be independent 
of the system size. In Fig. \ref{fig2} we plot the behavior
of $\gamma$ as function of $W$ for several sizes of a scale free graph.
Indeed, $\gamma$ decreases 
with system size for small values of $W$ while it 
increases with size for large
values of $W$.

\begin{figure}\centering
\epsfxsize6cm\epsfbox{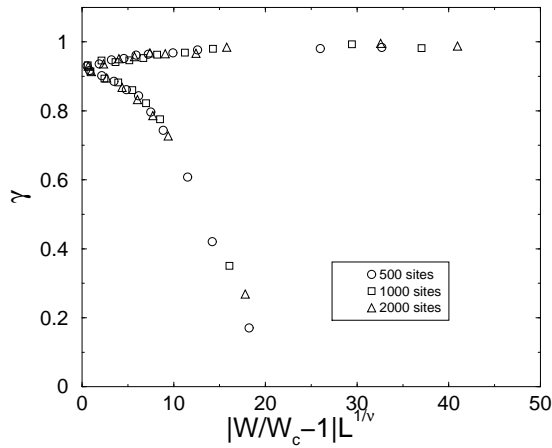} \caption{ The scaling of
$\gamma$ according to Eq. (\ref{scaling}) for different SF, $\lambda=4,m=2$, networks sizes.
Two branches, corresponding to the metallic and localized
regimes, appear.} \label{fig3}
\end{figure}

All curves should cross around a particular value of disorder signifying 
the critical disorder.
From finite size scaling arguments \cite{shklov92} 
one expects that $\gamma$ around the critical disorder will depend on the the
disorder and network size, $L$, in the following way:
\begin{eqnarray}
\gamma(W,L) = \gamma(W_c,L) + C\left|{{W}\over{W_c}}
-1\right|L^{1/\nu} , \label{scaling}
\end{eqnarray}
where $C$ is a constant. This relation enables us to extract both
the critical disorder $W_c$ and the critical index $\nu$.
Scaling of the numerical data according to Eq. (\ref{scaling}) 
yields two branches corresponding to the metallic and localized
regimes, that are clearly seen in Fig. \ref{fig3}. 
The estimated values of $\nu$ and $W_c$ 
(see Table \rm{I})
are extracted by fitting the branches 
to a 4th order polynomial.

\section{Results and Discussion}
\label{s3}

The calculations for all networks mentioned above are performed 
for $M$ different realizations, where
$M = 1000, 400, 200, \ldots , 50$ for the corresponding number of nodes: 
$N = 200, 500 ,1000, \ldots , 4000$.
Except for the Cayley-tree  networks for which
$M = 4000, 2000, 1000, \ldots, 125, 64$ 
for the corresponding tree sizes: $N = 63, 127, 255,\ldots , 2047, 4095$
or $L = 6, 7, 8, \dots , 11, 12$ 
(where $L$ is the number of "generations" of the tree).
Another exception is for Erd\"os-R\'enyi graphs in which
$\langle k \rangle$ is between $3$ and $3.5$. 
The low connectivity of the graphs, results in one main 
cluster and relatively large number of not-connected nodes (about $5$\%). Thus, the calculations are made only for
the largest cluster of each realization, since a procedure that considers 
all the nodes is skewed by the eigenvalues of small disconnected clusters 
\cite{berkovits96}.
A clear localization transition is observed for a group of graphs 
which are all characterized by an average degree
$\langle k \rangle$ smaller than $3.1$, and an averaged last
occupied shell $l$ (for $N=1000$ sites) larger or equal to $9.45$.
The results are summarized in Table {\rm I}.

\begin{table} \label{t1}
{\protect
\begin{tabular}{|c|c|c|c|c|}
\hline
Network  & $\langle k \rangle$ & $l$ & $W_c$ & $\nu$ \\ \hline
Scale-Free , $\lambda=4$ , $m=2$ & $2.97$ & $12.46$ & $15.7{\pm 0.9}$ & $0.55{\pm 0.11}$\\ \hline
Random-Regular (RR)& $3$ & $11.8$ & $11.9{\pm 0.26}$ & $0.66{\pm 0.08}$\\ \hline
Erd\"os-R\'enyi & $3$ & $9.45$ & $20.5{\pm 0.23}$ & $0.68{\pm 0.08}$\\ \hline
Cayley-Tree & $3$ & $10$ & $11.44{\pm 0.06}$ & $0.51{\pm 0.045}$\\ \hline
Cayley-Tree with loops& $3.05$ & $10$ & $12.4{\pm 0.1}$ & $0.54{\pm 0.075}$ \\ \hline
RR "double peak" & $3.1$ & $10.28$ & $14.1{\pm 0.3}$ & $0.85{\pm 0.41}$ \\ \hline
\end{tabular}}
\caption{\small Networks showing the localization transition. The value of $l$
is for $N=1000$.}
\end{table}

\begin{figure}\centering
\epsfxsize6cm\emph{}\epsfbox{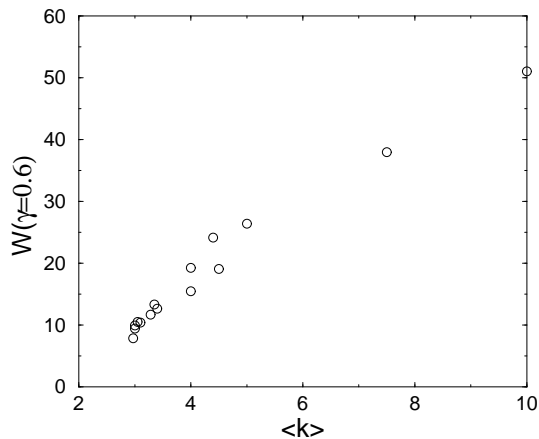} \caption{ The $\gamma$ curves 
of the checked networks can be ordered by their $\langle k \rangle$.
The values of their $W(\gamma=0.6)$ are presented as a function of 
$\langle k \rangle$. One can notice the increasing $W$
with $\langle k \rangle$.} \label{fig4}
\end{figure}

The results for all the graphs (including those which show no 
clear signs of transition) 
can be scaled according to their average degree $\langle k \rangle$. 
The higher the value of $\langle k \rangle$ is - 
the higher is the value of $W$ needed in order
to obtain a specific value of $\gamma$.
Thus, the higher the average degree, the more metallic
the system is, which makes sense.
A cross section at $\gamma=0.6$ of all curves is shown
in Fig.\ref{fig4} as a function of $\langle k \rangle$.
The $\langle k \rangle$ of the networks studied in Fig. \ref{fig4} as well as the averaged last
occupied shell $l$ for $N=1000$ sites
are presented in Table {\rm II}. 


\begin{table} \label{t2}
{\protect
\begin{tabular}{|c|c|c|}
\hline
 Network  & $\langle k \rangle$ & $l$ (for $N=1000$)\\ \hline \hline
Scale-Free , $\lambda=4$ , $m=2$ & $2.97$ & $12.46$ \\ \hline
Random-Regular & $3$ & $11.8$ \\ \hline
Random-Regular "double peak"  & &\\ ($p=0.95 \rightarrow k=3$ , $p=0.05 \rightarrow k=5$) & $3.1$ & $10.28$ \\ \hline
Cayley-Tree & $3$ & $10$ \\ \hline
Cayley-Tree with loops&  $3.05$ & $10$\\ \hline
Erd\"os-R\'enyi& $3$ & $9.45$ \\ \hline
Scale-Free , $\lambda=3.5$ , $m=2$ & $3.28$ & $9.36$ \\ \hline
Erd\"os-R\'enyi& $3.1$ & $9.31$ \\ \hline
Erd\"os-R\'enyi& $3.2$ & $9.03$ \\ \hline
Random-Regular "double peak"   & & \\ ($p=0.8 \rightarrow k=3$ , $p=0.2 \rightarrow k=5$) &  $ 3.4 $ & $9$  \\ \hline
Erd\"os-R\'enyi& $3.5$ & $8.33$ \\ \hline
Random-Regular "double peak"  & & \\ ($p=0.95 \rightarrow k=3$ , $p=0.05 \rightarrow k=10$)& $3.35$ & $7.99$ \\ \hline
Erd\"os-R\'enyi& $4$ & $7.51$ \\ \hline
Scale-Free , $\lambda=5$ , $m=3$ & $4$ & $7.37$ \\ \hline
Scale-Free , $\lambda=4$ , $m=3$ & $4.5$ & $6.05$\\ \hline
Random-Regular "double peak"  & &\\ ($p=0.8 \rightarrow k=3$ , $p=0.2 \rightarrow k=10$)& $4.4$ & $6.13$ \\ \hline
Erd\"os-R\'enyi& $5$ & $6.31$ \\ \hline
Erd\"os-R\'enyi& $7.5$ & $5.02$ \\ \hline
Erd\"os-R\'enyi& $10$ & $4.1$ \\ \hline
\end{tabular}}
\caption{The average connectivity $\langle k \rangle$ of all the networks 
considered in this study, as well as the averaged last
occupied shell $l$ for $N=1000$ sites.}
\end{table}

The following observations can be gleaned out of the data for the different
networks:\\
(1) For all the networks that show a metal-insulator transition,
$\nu$ is of order $1/2$ except for the Random-Regular "double-peak" network
which is the one with the highest value of connectivity that still shows
a clear transition. A critical index of $\nu=0.5$ is expected for a
system of infinite dimensionality. At $\langle k \rangle = 3.1$ the value of $\nu$ is significantly
higher, but so is the estimate of the error bar. On the other hand for the Erd\"os-R\'enyi
graph with $\langle k \rangle = 3.1$ no clear transition is observed.\\
(2) All networks with connectivity above $3.1$ do not show clear
signs of a metal-insulator transition. Nevertheless, one should be 
rather careful in interpreting this observation since, as is clear
from Table {\rm II}, larger values of  $\langle k \rangle$  lead
to smaller size, $l$, of the network for the same number of nodes. 
Moreover, from the two networks which have the same $\langle k \rangle=3.1$,
only the one with the higher value of $l$ shows clear signs of the metal insulator transition.
Thus, the absence of transition may be an artifact of the small size
of networks with high average connectivity.\\
(3) The critical disorder $W_c$ fluctuates in the range of
$12-20$ (Table \rm{I}). 
Due to the small range of $\langle k \rangle$ ($2.97-3.1$), 
it is hard to determine any relation between $k$ and $W_c$.\\
(4) On the other hand, there is a clear relation between
the amount of disorder needed in order to reach a particular
value of $\gamma$ (i.e., the value of $W$ needed to reach a certain
degree of localization) and  $\langle k \rangle$. As can be seen in
 Fig. \ref{fig4}, a linear dependence $W(\gamma=0.6)\propto \langle k \rangle$
is observed.

Thus, the gross features of the Anderson metal-insulator transition
are similar for a wide range of different networks. The critical
index for all the networks studied here are within the range expected for
a system of infinite dimensionality, and the connectivity influences
the degree of localization. 
Thus, complex networks are an example of a topology for which the critical
index  follows the mean field prediction, but the critical disorder
can be tuned by the connectivity. 
On the other hand, the fact that networks with
high connectivity are very compact raises the problem of
identifying  the transition point. It is hard to extend
the usual finite-size scaling method to networks with high connectivity
since the number of sites grows very rapidly with size, while for
small network sizes the crossover behavior of the $\gamma$ curves
is very noisy. This results in an inability to clearly identify
the Anderson transition, although it can not be ruled out the 
possibility that there
is a critical connectivity for complex networks above which no transition
exist.

We acknowledge very useful discussions with J. W. Kantelhardt and helpful
correspondence with Y.V. Fyodorov, as well
as support from the Israel Academy of Science (Grant 877/04), the ONR, and the
Israeli Center for Complexity Science. This work was also supported by
the European research NEST Project No. DYSONET 012911.

\end{document}